\documentstyle{ichep}
\input epsf
\global\arraycolsep=1.8pt
\def\sss{\scriptscriptstyle}
\def\barp{{\raise.35ex\hbox{${\sss (}$}}---{\raise.35ex\hbox{${\sss
)}$}}}
\def\bdbarp{\hbox{$B_d$\kern-1.4em\raise1.4ex\hbox{\barp}}}
\def\bsbarp{\hbox{$B_s$\kern-1.4em\raise1.4ex\hbox{\barp}}}

\newcommand{\beq}{\begin{equation}}
\newcommand{\eeq}{\end{equation}}
\newcommand{\absvcb}{\vert V_{cb}\vert}

\def\mb{m_b}

\newcommand{\bgamaxs}{$B \to X _{s} \gamma$}

\newcommand{\BBGAMAXS}{{\cal B} (B \to X _{s} \gamma)}
\newcommand{\brbgamaxs}{${\cal B} (B \to X _{s} \gamma)$}
\newcommand{\brbksgam}{${\cal B}(B \to K^* \gamma)$}
\newcommand{\bksgam}{ $B \to K^* \gamma$}

\newcommand{\Bsell}
   {$B \to X_s ~\ell^+ \ell^-$}

\def\qbar{\overline q}

\def\q5q{\qbar{{\lambda_a}\over 2} i\gamma_5 q}

\def\to{\rightarrow}

\def\mb{m_b}

\def\s{\hat s}

\begin{document}
%---------------- CERN Titlepage <---------------------------
\thispagestyle{empty}
%\begin{titlepage}

\begin{flushright}
CERN-TH.7430/94
\end{flushright}

\vspace{0.3cm}

\begin{center}
\Large\bf Towards a Model Independent Analysis of Rare $B$ Decays  \\
\end{center}

\vspace{0.8cm}

\begin{center}
Ahmed Ali, Gian Giudice and Thomas Mannel  \\
{\sl Theory Division, CERN, CH-1211 Geneva 23, Switzerland}
\end{center}

\vspace{0.8cm}

\begin{abstract}
\noindent
We propose to undertake
a model-independent analysis of the inclusive decay rates and
distributions in the processes \bgamaxs~ and  \Bsell ~($B=B^\pm$ or
$B^0_d$). We show how measurements
of the decay rates and distributions in these processes
would allow us to extract the magnitude and sign of the dominant
Wilson coefficients of the magnetic moment operator
$\mb \bar{s}_L \sigma_{\mu \nu} b_R F^{\mu \nu }$ and the four-fermion
operators $(\bar{s}_L \gamma_\mu b_L)(\bar{\ell} \gamma^{\mu} \ell)$
and $(\bar{s}_L \gamma_\mu b_L)(\bar{\ell} \gamma^{\mu}\gamma^5 \ell)$.
\end{abstract}
\vfill
\begin{center}
Contribution to the \\
            $27^{th}$ {\it International Conference on
            High Energy Physics } \\
            Glasgow, Scottland, 20 -- 27 July 1994
\end{center}
\bigskip\bigskip
\vfill
\noindent
CERN-TH.7430/94 \\
October 1994
%\end{titlepage}
%---------------- END CERN Titlepage <---------------------------
\boldmath
\title{Towards a Model Independent Analysis of Rare $B$ Decays}
\unboldmath

\author{Ahmed Ali$^{\ddag\P}$\ , Gian Giudice$^{\S\dag}$\ and
  Thomas Mannel$^{\|}$}

\affil{Theory Division, CERN, CH-1211 Geneva 23, Switzerland.}

\abstract{
We propose to undertake
a model-independent analysis of the inclusive decay rates and
distributions in the processes \bgamaxs~ and  \Bsell ~($B=B^\pm$ or
$B^0_d$). We show how measurements
of the decay rates and distributions in these processes
would allow us to extract the magnitude and sign of the dominant
Wilson coefficients of the magnetic moment operator
$\mb \bar{s}_L \sigma_{\mu \nu} b_R F^{\mu \nu }$ and the four-fermion
operators $(\bar{s}_L \gamma_\mu b_L)(\bar{\ell} \gamma^{\mu} \ell)$
and $(\bar{s}_L \gamma_\mu b_L)(\bar{\ell} \gamma^{\mu}\gamma^5 \ell)$.}

\twocolumn[\maketitle]

\fnm{2}{e-mail: alia@cernvm.cern.ch}
\fnm{5}{On leave of absence from DESY, Hamburg, FRG.}
\fnm{3}{e-mail: giudice@vxcern.cern.ch}
\fnm{1}{On leave of absence from INFN, Sezione de Padova, Italy.}
\fnm{4}{e-mail: mannel@cernvm.cern.ch}

\boldmath
\section{ The Decay \bgamaxs ~in SM and Experiment}
\unboldmath
The measurements of the decay mode \bksgam , reported last year
by the CLEO collaboration
\cite{CLEOrare1}, having a branching ratio
\brbksgam ~$=(4.5 \pm 1.0 \pm 0.9)\times 10^{-5}$, and
the inclusive
decay $B \to X_s \gamma$,
reported at this conference
\cite{CLEObsg} with a branching ratio $
{\cal B} (B \to X_s \gamma ) = (2.32 \pm 0.51 \pm 0.32 \pm 0.20)
 \times 10^{-4}$,
have put the physics
of the electromagnetic penguins on an experimental footing.
In the standard model (SM), these transitions are dominated by
the short-distance
contributions and
provide valuable information about the top quark mass and
the Cabibbo-Kobayashi-Maskawa (CKM) weak mixing matrix elements
$V_{ts}V_{tb}$ \cite{CKM}. The rapport between the SM and experiment
 may be quantified in terms of the CKM matrix element ratio
 \cite{ag5}:
\beq
0.62 \leq \left\vert {V_{ts} \over V_{cb}} \right\vert \leq 1.1~,
\label{vtslim}
\eeq
which is consistent with unity, resulting from the unitarity constraints.
Alternatively, one can set
$V_{ts} / V_{cb} =1$
to obtain from the CLEO measurement bounds on the
Wilson coefficient $C_7(\mb)$ of the effective magnetic moment operator.
Using ${\cal B} (B \to X_s \gamma )$ from \cite{CLEObsg}, one obtains
\beq
0.22 \leq \vert C_7(\mb) \vert \leq 0.30.
\label{c7bound}
\eeq
Using, however, the 90\%-confidence-level range from the CLEO measurement
$\BBGAMAXS = (2.31 \pm 1.1)  \times 10^{-4}$ and the theoretical
calculation for $\BBGAMAXS$ from \cite{ag5} we obtain
\begin{equation}
  0.19 \leq \vert C_7(\mb) \vert \leq 0.32.
\end{equation}
We also remark that the photon energy and hadron mass spectra measured
by CLEO are in good agreement with the SM-based calculations in
\cite{ag1}.
The bound (\ref{c7bound}) can be
used to constrain the non-SM contribution to the decay rate \brbgamaxs
{}~as discussed in these proceedings \cite{desh94,nath94}.

%%%%%%%%%%%%%%%%%%%%%%%%%%%%%%%%%%%%%%%%%%%%%%%%%%%%%%%%%%%%%%%%%%%%%%%%

% Although the method of analysing
%the data and the relevant theoretical framework in ref. ~\cite{AGMSusy}
%are developed for the inclusive decays
%\bgamaxs ~and $B \to X_s + \ell^+ \ell^-$,
%much of the general considerations being discussed apply also to
%the corresponding exclusive decays such as \bksgam, ~\bksell ~and
%$B \to K \ell^+ \ell^-$. Of course, the extraction of the
%short-distance physics in
%terms of the Wilson coefficients of the dominant operators from the
%data on exclusive decays would require the knowledge of the relevant
%form factors.

%%%%%%%%%%%%%%%%%%%%%% SECTION 2 %%%%%%%%%%%%%%%%%%%%%%%%%%%%%%%%
\boldmath
\section{Motivation for a Model Independent Analysis of Rare $B$ Decays}
\unboldmath
The determination of $\vert C_7(\mb)\vert$ from the inclusive branching
ratio \brbgamaxs ~is a prototype of the kind of analysis that we
would like to propose here to be carried out for the rare $B$ decays
in general and for the semileptonic decays \Bsell , in particular.
First steps towards a
model-independent analysis of the FCNC electroweak
rare $B$ decays involving these decay modes
have recently been proposed in \cite{AGMSusy}, to which we refer
for details and references to other related work. Here, we summarize the
main assumptions and results.

The main interest in rare $B$ decays is to measure the effective FCNC
vertices in order to test the SM and  search for new physics.
We have argued that with some plausible assumptions
  these vertices can be parametrized through a
limited number of effective parameters, which govern the rates and
shapes (differential
distributions) in rare $B$ decays $B \to X_s \gamma$, \Bsell and
$B_s \to \ell^+ \ell^-$. The search for
physics beyond the SM in these decays can be carried out in terms of
three effective parameters, $C_7(\mu), ~C_9(\mu)$ and $C_{10}(\mu)$,
characterizing the strength of the magnetic moment and two four-fermion
operators $(\bar{s}_L \gamma_\mu b_L)(\bar{\ell} \gamma^{\mu} \ell)$
and $(\bar{s}_L \gamma_\mu b_L)(\bar{\ell} \gamma^{\mu}\gamma^5 \ell)$.
 This can then be interpreted in a large
class of models. The presence of non-SM physics may manifest itself by
distorting the differential distributions in \Bsell . Some possible
examples of such distortions have been worked out
in \cite{AGMSusy}. Here we present
profiles of the Wilson coefficients in the best-motivated
extensions of the SM, namely the Minimal Supersymmetric Standard Model
(MSSM).

Our analysis is based on an
effective Hamiltonian of the form
\beq
{\cal H}_{eff}(b \to sX) = -\frac{4G_F}{\sqrt{2}} \lambda_t
            \sum_{i=1}^{10} C_i(\mu) {\cal O}_i(\mu) ~.
\label{heff}
\eeq
where $X$ stands for $q\bar{q}, ~\gamma$, gluon and $\ell^+ \ell^-$ and
$\lambda_t = V_{ts}^*V_{tb}$. The operator basis ${\cal O}_{1 \cdots 10}$
is given in \cite{AGMSusy} and is the same as in the SM, thereby restricting
our analysis to cases, in which the effective Hamiltonian may be written
as (\ref{heff}).

\boldmath
\section{Analysis of the Decays $B \to X_s \gamma$ and
         $B \to X_s \ell^+ \ell^-$}
\unboldmath
The experimental quantities we consider in this paper are the
following: \boldmath {\bf
(i) Inclusive radiative rare decay branching ratio \brbgamaxs ;
(ii) Invariant dilepton mass distributions in \Bsell ;
(iii) Forward-backward (FB) charge asymmetry
 ${\cal A}(\hat{s})$ in \Bsell}\unboldmath .

The FB asymmetry
${\cal A}(\hat{s})$ is defined with respect to the angular variable
$z \equiv \cos \theta $, where $\theta$ is the angle of the
$\ell^+$ with respect to the $b$-quark direction
in the centre-of-mass system of
the dilepton pair.
It is obtained by integrating the doubly differential
distribution $d^2 {\cal B} / (dz \, d\hat{s})$(\Bsell ) \cite{AMM}:
\beq \label{asy}
{\cal A}(\hat{s}) \equiv \int\limits_0^1 dz \,
\frac{d^2{\cal B}}{dz \, d\hat{s}}  -
\int\limits_{-1}^0 dz \,
\frac{d^2{\cal B}}{dz \, d\hat{s}} ,
\eeq
where $\hat{s} = (p_1 + p_2)^2/\mb^2$ and
$p_1$ and $p_2$ denote, respectively, the momenta of the
 $\ell^+$ and $\ell^-$.

We remark that the decay rate \brbgamaxs ~puts a bound
on the absolute value of the coefficient $C_7(\mu)$.
However, the radiative $B$ decay rate by itself
is not able to distinguish
between the solutions $C_7(\mu) > 0$ (holding in the SM) and the
solutions $C_7(\mu) < 0$, which, for example, are
also allowed in the MSSM
as one scans over the allowed parameter space. We recall that
the invariant dilepton mass distribution and the forward-backward asymmetry
in \Bsell ~are sensitive to the sign and magnitude of $C_7(\mu)$
\cite{AMM}.
Using ${\cal H}_{eff}$ given in
(\ref{heff}), one obtains for the dilepton invariant mass
distribution
\begin{eqnarray}
&& {d{\cal B} \over d\s}
 =  {\cal B}_{sl} \frac{\alpha^2}{4 \pi^2} \frac{\lambda_t}{\absvcb}^2
 \frac{\hat{w}(\hat{s})}{f(m_c/m_b)}
\\ \nonumber
&& \times \left[ \vphantom{\frac{1}{1}}
\left( |C_9 + Y(\hat{s})|^2 + C_{10}^2 \right)
\alpha_1 (\hat{s},\hat{m}_s) \right.
\\ \nonumber
&& \quad \left. + \frac{4}{\hat{s}} C_7^2 \alpha_2 (\hat{s},\hat{m}_s)
+ 12 \alpha_3 (\hat{s},\hat{m}_s) C_7
(C_9 + \mbox{ Re }Y(s)) \right] ,
\label{dbrs}
\end{eqnarray}
where the auxiliary functions $\alpha_i$ depend only on the
kinematic variables and $Y(\hat{s})$ depends on the coefficients
$C_1,\cdots,C_6$ of the four quark operators (see \cite{AGMSusy}).

The corresponding differential asymmetry as defined in (\ref{asy}) is
\begin{eqnarray}
{\cal A} (\hat{s}) &=& - {\cal B}_{sl} \frac{3 \alpha^2}{8 \pi^2}
                      \frac{1}{f(m_c/m_b)} \hat{w}^2 (\hat{s})
C_{10}  \\ \nonumber
&&  \times \left[ \hat{s} ( C_9 + \mbox{ Re } Y(\hat{s})) +
4 C_7 (1 + \hat{m}_s^2) \right] .
\label{dasym}
\end{eqnarray}

We first present  the partial branching
ratio ${\cal B}(\Delta  s ) $ and partial FB asymmetry
${\cal A} (\Delta  s ) $,
where $\Delta  s $ defines an interval in the dilepton invariant
mass.
In order to minimize long-distance effects we shall
consider the kinematic regime for $s$ below the $J/\psi$ mass
(low invariant mass) and for $s$ above the mass of the $\psi '$
(high invariant mass). Integrating (\ref{dbrs}) over these regions for
the invariant mass one finds
\begin{equation} \label{branch}
{\cal B} (\Delta s) = A(\Delta s) \left( C_9^2 + C_{10}^2 \right)
                      + B(\Delta s) C_9 + C(\Delta s) ,
\label{bds}
\end{equation}
where $A$, $B$ and $C$ are fixed in terms of the Wilson coefficients
$C_1 \cdots C_6$ and $C_7$.
 For the numerical analysis we use $m_b = 4.7$ GeV,
$m_c = 1.5$ GeV, $m_s = 0.5$ GeV.
The resulting coefficients $A$, $B$, and $C$ are listed in Table
\ref{tab1} for the decays
 $B \to X_s e^+ e^-$ and $B \to X_s \mu^+ \mu^-$.

\begin{table}
\begin{center}
\begin{tabular}{| c | c | c | c | c | c |}
\hline
 $\Delta s$ & $C_7$   & $A(\Delta s)$
                              & $B(\Delta s)$
                              & $C(\Delta s)$
                              & $C(\Delta s)$  \\
 & & & & $\ell = e$ & $\ell = \mu$ \\
\hline \hline
$4m_\ell^2 < s < m_{J/\psi}^2$& $+0.3$ & 2.86 & $-$5.76 & 84.1 & 76.6 \\
$4m_\ell^2 < s < m_{J/\psi}^2$& $-0.3$ & 2.86 & $-$20.8 & 124  &  116 \\
$m_{\psi'}^2 < s < (1-m_s^2)$& $+0.3$ & 0.224 & $-$0.715 & 0.654 & 0.654 \\
$m_{\psi'}^2 < s < (1-m_s^2)$& $-0.3$ & 0.224 & $-$1.34  & 2.32  & 2.32 \\
\hline
\end{tabular}
\end{center}
\caption{Values for the coefficients $A(\Delta s)$, $B(\Delta s)$
         and $C(\Delta s)$ (in units of $10^{-8}$) for the decay
         $B \to X_s \ell^+ \ell^-$.
         }
\label{tab1}
\end{table}

For a measured branching fraction ${\cal B} (\Delta s)$, one can solve
the above equation for ${\cal B }(\Delta s)$, obtaining a
circle  in the $C_9$-$C_{10}$ plane, with  centre lying at
$C_9^* =  B(\Delta s)/(2 A(\Delta s))$ and $ C_{10}^* = 0$.
The radius  of this circle is proportional to
$
\sqrt{{\cal B}(\Delta s)-{\cal B}_{min}(\Delta s)} ,
$
where the minimum branching fraction
\begin{equation}
{\cal B}_{min}(\Delta s) = C(\Delta s) -
            \frac{B^2(\Delta s)}{4 A(\Delta s)}
\end{equation}
is determined mainly by the present data on $B \to X_s \gamma$,
i.e.~by $|C_7|$.

To further pin down the Wilson coefficients,
one could perform a measurement of the forward-backward
asymmetry ${\cal A}$.  Integrating (\ref{asy})
over a range $(\Delta s)$ yields
\begin{equation} \label{FB}
{\cal A} (\Delta s) = C_{10}
\left(\alpha (\Delta  s) C_9 + \beta(\Delta  s) \right) .
\end{equation}

For a fixed value of ${\cal A} (\Delta s)$, one obtains
hyperbolic curves in the $C_9 $-$C_{10}$ plane; like
the coefficients $A$, $B$ and $C$, the parameters $\alpha$ and $\beta$ are
given in terms of the Wilson coefficients $C_1 \cdots C_6$, $C_7$
and $\Delta s$; their values are presented in
Table \ref{tab2}.

\begin{table}
\begin{center}
\begin{tabular}{| c | c | c | c | }
\hline
 $\Delta s$ & $C_7$ & $\alpha(\Delta s) $
                            & $\beta (\Delta s) $ \\
\hline \hline
$4m_\ell^2 < s < m_{J/\psi}^2$& $+0.3$ & $-$6.08 & $-$24.0  \\
$4m_\ell^2 < s < m_{J/\psi}^2$& $-0.3$ & $-$6.08 &   55.4   \\
$m_{\psi'}^2 < s < (1-m_s^2)$& $+0.3$ & $-$0.391 & 0.276 \\
$m_{\psi'}^2 < s < (1-m_s^2)$& $-0.3$ & $-$0.391 & 1.37  \\
\hline
\end{tabular}
\end{center}
\caption{Values for the coefficients $\alpha(\Delta s)$and
         $\beta(\Delta s)$ ( in units of $10^{-9}$).}
\label{tab2}
\end{table}

Given the two experimental inputs, the branching fraction
${\cal B}(\Delta s)$ and the corresponding asymmetry ${\cal A}(\Delta s)$,
one obtains a fourth-order equation for the Wilson coefficients
$C_9$ and $C_{10}$, which admits in general four solutions, which can
be plotted as contours for a fixed
value for the branching fraction ${\cal B} (\Delta s)$ and the
FB asymmetry
${\cal A} (\Delta  s)$.
The possible solutions
for $C_9$ and $C_{10}$ are given by the intersections of the circle
corresponding to the measured branching fraction and the hyperbola,
corresponding to the measured asymmetry.
Details can be seen in \cite{AGMSusy}.

\begin{figure}
%   \vspace{-0.5cm}
   \epsfysize=5cm
   \centerline{\epsffile{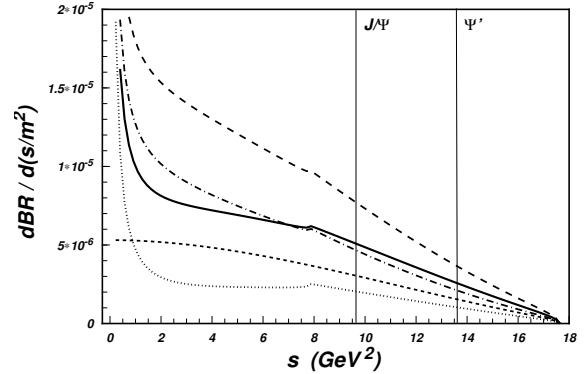}}
   \caption{The dependence of the invariant-mass spectrum on the Wilson
coefficients.
Solid line: SM. Long-dashed line: $C_7 \to - C_7$, with other
coefficients retaining their SM values.
Short-dashed line: The contribution of $C_{10}$ only.
Dotted line: $C_{10} = 0$, with other coefficients retaining their SM values.
Dash-dotted line:  same as for the dotted one, but with $C_7 = -0.3$.
The vertical lines indicate
the location of the $J/\Psi$ and $\Psi^\prime$ resonances. }
\label{fig5}
\end{figure}

We stress that the spectrum
itself is very sensitive to the values of the Wilson coefficients
and to the sign of $C_7$. In fig.~\ref{fig5} we plot the various
contributions to the spectrum, for positive and for negative $C_7$.

\begin{figure}
%  \vspace{-0.5cm}
   \epsfysize=5cm
   \centerline{\epsffile{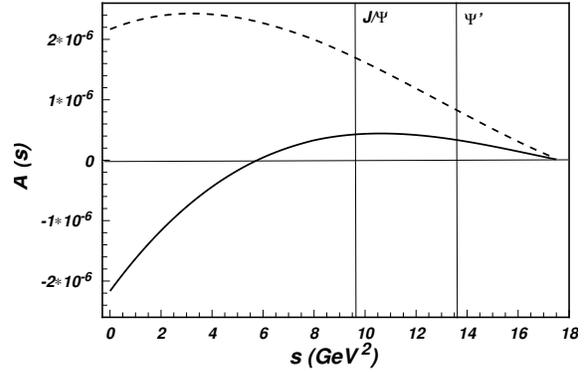}}
   \caption{The dependence of the differential FB asymmetry on the Wilson
coefficients.
Solid line: SM. Long-dashed line: $C_7 \to - C_7$, with the other
parameters retaining their SM values.
The vertical lines indicate
the location of the $J/\Psi$ and $\Psi^\prime$ resonances.}
\label{fig6}
\end{figure}

In a similar way, it may become possible to measure also the
differential asymmetry (\ref{asy}); the various
contributions to ${\cal A} (s)$ are shown in fig.~\ref{fig6}.

%%%%%%%%%%%%%%%%%%%%%% SECTION 4 %%%%%%%%%%%%%%%%%%%%%%%%%%%%%%%%

\section{Model Predictions for the Wilson Coefficients}

As an illustrative example we shall consider here the MSSM;
we shall show, how the predictions for
the Wilson coefficients $C_7$, $C_9$, and $C_{10}$ are altered in
this model.

Once the gluino contributions (as well as the analogous ones from
neutralino exchange) are neglected, the flavour violation in the
supersymmetric models is completely specified by the familiar
CKM matrix. The one-loop supersymmetric corrections
to the Wilson coefficients
$C_7$, $C_9$, and $C_{10}$ are given by two classes of diagrams:
charged-Higgs exchange and chargino exchange  \cite{bbmr}.

The charged-Higgs contribution is specified by two input parameters:
the charged-Higgs mass ($m_{H^+}$) and the ratio of Higgs vacuum
expectation values ($v_2/v_1\equiv \tan \beta$). This
contribution alone corresponds to the two-Higgs doublet model
which has also been considered in \cite{AGMSusy}.

In addition to the diagrams with charged-Higgs exchange, the MSSM
leads also to chargino-mediated diagrams.
The chargino contribution is specified by six parameters.
Three of them enter the $2\times 2$ chargino mass matrix:
\begin{equation}
m_{\chi^+}=\pmatrix{M & m_W\sqrt{2}\sin \beta \cr
 m_W\sqrt{2}\cos \beta & \mu }.
\end{equation}
Following standard notations, we call $\tan \beta$ the ratio of
vacuum expectation values, the
same that appears also in the charged-Higgs sector, and
$M$, $\mu$ the gaugino and higgsino mass parameters,
subject to the constraint that the lightest chargino mass satisfies
the LEP bound, $m_\chi^+>45$ GeV. The squark masses
\begin{equation}
m^2_{{\tilde{q}}^2_\pm}={\widetilde{m}}^2+m_q^2\pm A{\widetilde{m}}m_q
\label{sq}
\end{equation}
contain two additional free parameters besides the known mass of
the corresponding quark $m_q$: a common supersymmetry-breaking mass
$\widetilde{m}$ and the coefficient $A$.
The last parameter included in our analysis is a common mass $m_{\tilde{l}}$
for sleptons, all taken to be
degenerate in mass, with the constraint $m_{\tilde{l}}> 45$ GeV.
Therefore the version of the MSSM we are considering is defined in terms
of seven free parameters.

\begin{figure}
%   \vspace{-0.3cm}
   \epsfysize=4.8cm
   \centerline{\epsffile{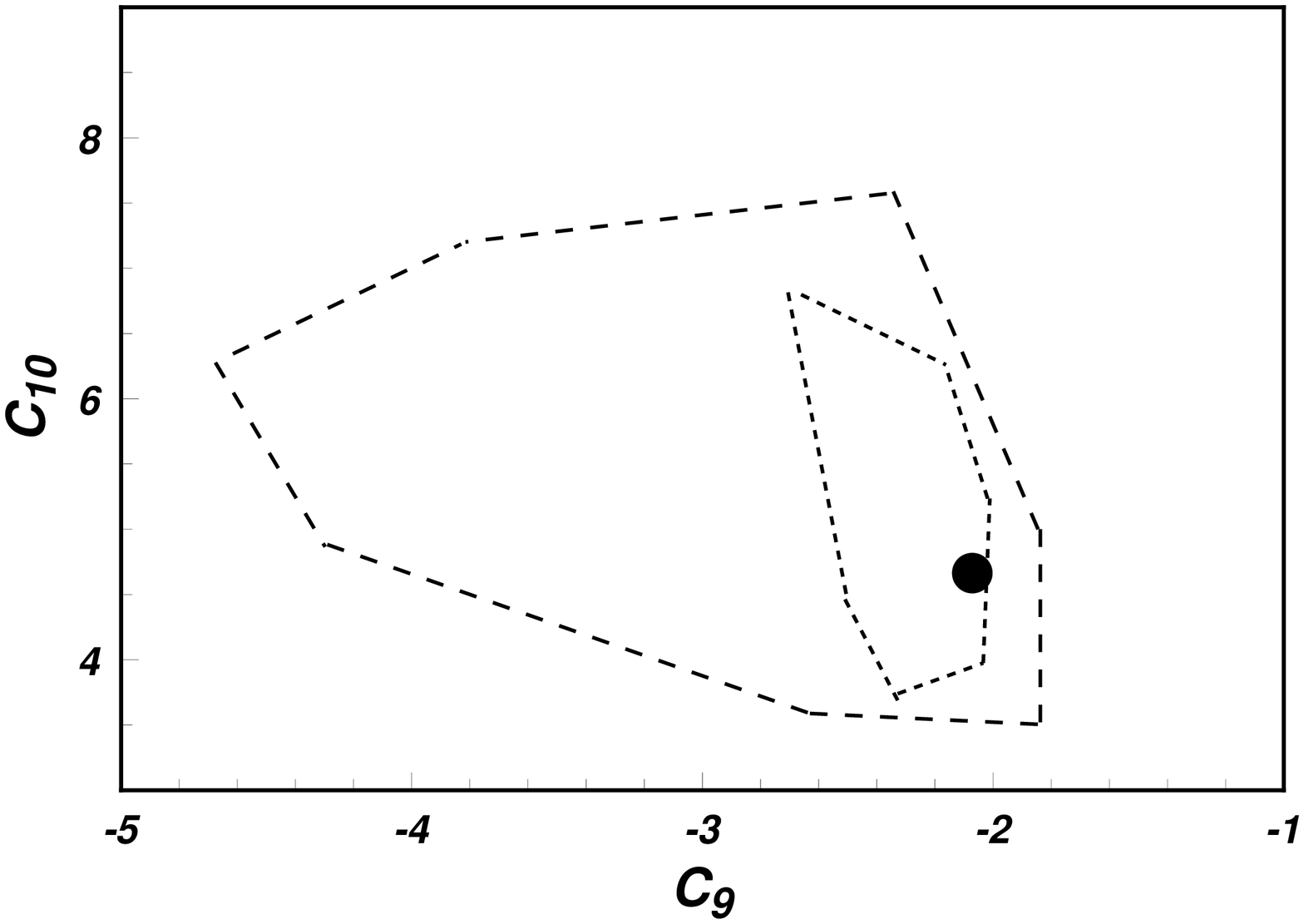}}
   \epsfysize=4.8cm
   \centerline{\epsffile{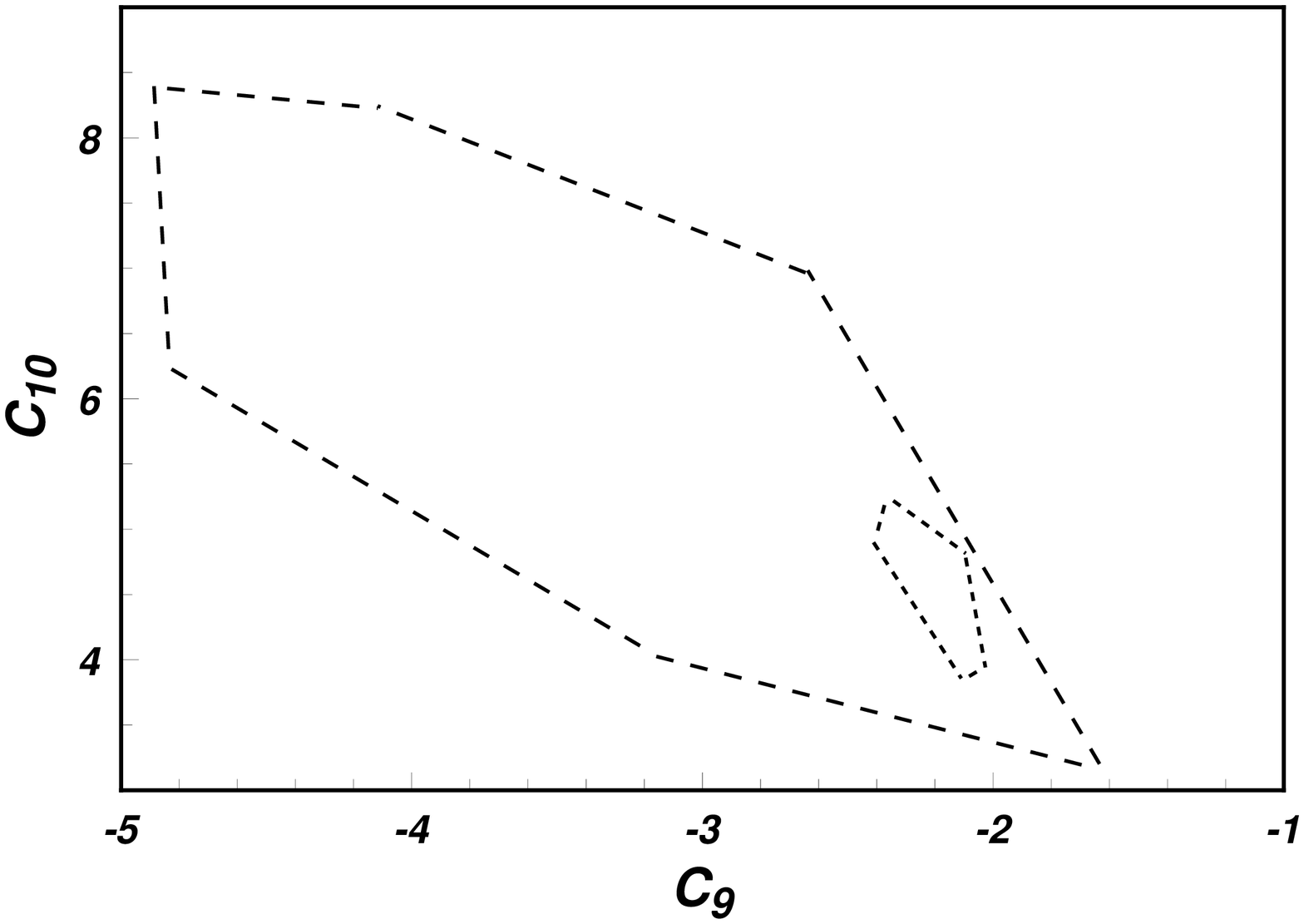}}
   \caption{The region in the $C_9$-$C_{10}$ plane
obtained by varying the MSSM parameters.
The upper (lower) plot corresponds to solutions that satisfy
the $b\to s\gamma$ experimental constraint with positive (negative)
$C_7$ given in eq.(\protect{\ref{c7bound}}) and the present bounds
($m_{H^+} > $ 80 GeV, $\tilde{m}_t, m_{\chi^+}, \tilde{m}_\ell > $ 45 GeV).
The smaller areas limited by the short-dashed line correspond to the
region of the MSSM parameter space that will survive an unsuccessful
search for supersymmetry at the Tevatron and LEP 200 ($m_{H^+}>150$ GeV,
$m_{\tilde{t}},~ m_{\chi^+},~m_{\tilde{l}}>100$ GeV).}
\label{susy1}
\end{figure}

We have computed the Wilson coefficients in the MSSM
and then varied the seven above-defined parameters in
the experimentally allowed region.
The results of our analysis are presented in fig.~\ref{susy1},
which shows the regions of the $C_9$--$C_{10}$ plane allowed by
possible choices of the MSSM parameters. The upper plot of
fig.~\ref{susy1} corresponds
to parameters which give rise to positive (same sign as in the SM)
values of $C_7$, consistent with experimental results on $b\to
s\gamma$ ($0.19<C_7<0.32$), while the lower plot
corresponds to
values of $C_7$ with opposite sign ($-0.32<C_7<-0.19$). We also
show how our results are affected
by an improvement in
the experimental limits on supersymmetric particle
masses, as can be expected from the Tevatron and LEP 200.
Fig.~\ref{susy1} also shows the $C_9-C_{10}$ regions allowed by
the MSSM if the further
constraints $m_{H^+}>150$ GeV, $m_{\tilde{t}},~ m_{\chi^+}
,~m_{\tilde{l}}>100$ GeV are imposed.

The regions shown in fig.~\ref{susy1} illustrate the
typical trend of the supersymmetric corrections. If supersymmetric
particles exist at low energies, we can expect larger values of $C_{10}$
and smaller (negative) values
of $C_9$ than those predicted by the SM.
This is the general feature, although the exact boundaries
of the allowed regions depend on the particular model-dependent
assumptions one prefers to use.
 However, the most interesting feature of
supersymmetry is that solutions with negative values of $C_7$ are possible
and are still consistent with present data. Moreover, values of the
other two coefficients $C_9$ and $C_{10}$ sufficiently different from
the SM are allowed, leading to measurable differences in the decay rates
and distributions of $B \to X_s \ell^+ \ell^-$ and $B_s \to \ell^+ \ell^-$.

%\vskip0.5truein

\end{document}